\documentclass[aps,pra,preprint,showpacs,showkeys]{revtex4}
\usepackage{amsmath,amssymb}  
\usepackage{rotating,epsfig} 
\usepackage{bm}  
\usepackage{color}    
\begin{document}

\title{
Classical trajectory Monte Carlo model calculations for the antiproton-induced 
ionization of atomic hydrogen at low impact energy
}

\author{L. Sarkadi}
\email{sarkadil@atomki.hu}
\author{L. Guly\'as}
\affiliation{Institute for
Nuclear Research of the Hungarian Academy of Sciences
(ATOMKI), H-4001 Debrecen, Pf.\ 51, Hungary}

\date{\today}

\begin{abstract}
The three-body dynamics of the ionization of the atomic hydrogen by 30 keV
antiproton impact has been investigated by calculation of fully differential
cross sections (FDCS) using the classical trajectory Monte Carlo (CTMC) 
method. The results of the calculations are compared with the predictions of 
quantum  mechanical descriptions: The semi-classical time-dependent close-coupling 
theory, the fully quantal, time-independent close-coupling theory, and the 
continuum-distorted-wave-eikonal-initial-state model. In the analysis particular
emphasis was put on the role of the nucleus-nucleus (NN) interaction played in 
the ionization process. For low-energy electron ejection CTMC predicts a large
NN interaction effect on FDCS, in agreement with the quantum mechanical 
descriptions. By examining individual particle trajectories it was found that 
the relative motion between the electron and the nuclei is coupled very weakly 
with that between the nuclei, consequently the two motions can be treated
independently. A simple procedure is presented by which the NN interaction effect 
can be included into the calculations carried out without it.
\end{abstract}

\pacs{34.50.Fa, 34.10.+x}
\keywords{ionization, antiproton, fully differential cross section, CTMC}

\maketitle

\section{Introduction}

The ionization of the hydrogen atom by impact of antiprotons has attracted the 
attention of many theoreticians in the past decades. The great interest is 
explained by the fact that in the treatment of the process one is faced with a 
clean three-body break up problem: In contrast to proton impact there is no 
electron capture channel, and unlike the  electron-induced ionization, the 
treatment is not complicated by  electron exchange effects.

The enormous efforts devoted for the investigations of the collisions of 
antiprotons with atoms and molecules have been reviewed recently 
\cite{Kirchner11}. Besides the fundamental aspects of the topic, from the review 
the reader may learn about important, potential applications, for example, the
radiation therapy for cancer treatment. Although the dominant process utilized in 
the therapy is the annihilation, there are several aspects of atomic physics 
relevance of this application, e.g., the slowing down process of  the antiprotons 
in the biological issue, and the mechanism of creation of slow secondary electrons.

The subject of most of the research work carried out on the antiproton-induced
ionization of the hydrogen atom  was the energy-dependent total cross section. 
The few number of differential studies is explained partly by the present 
experimental limitations (first of all, the small intensity of the available
antiproton beam), and theoretically the difficulties arising in the calculations
of accurate partially or fully differential cross sections.

The deepest insight into the dynamics of the collision can be gained by
kinematically complete experiments. A technique used widely for this purpose in 
the field of atomic collisions is COLTRIMS (cold target recoil ion momentum
spectroscopy) \cite{Ullrich03}. COLTRIMS was applied in the only experimental
study in which differential cross sections were measured for collisions
involving antiprotons \cite{Khayat99}. In the experiment carried out for helium target
at 945 keV impact energy cross sections differential in the longitudinal 
electron and recoil-ion momenta were determined. The obtained data showed 
only a small ($<10\%$) difference from the corresponding cross sections 
measured by 1 MeV protons, as it is expected at such high impact energy.

The above experiment demonstrated the feasibility of differential measurements 
using antiprotons. This and future plans of facilities providing low-energy 
antiproton beams of high intensity (for a review see \cite{Siggel-King11}) gave 
great momentum to the theoretical investigations of the differential properties of 
the antiproton-induced ionization. Another motivation towards this direction was 
the clarification of the effect of nucleus-nucleus (NN) interaction on the fully
differential cross sections (FDCS) in ion-atom collisions. The role of the
NN interaction was one of the central questions of the attempts to solve the
long-standing puzzle regarding discrepancies between theory and experiment in
the FDCS for ionization in 100 MeV/amu C$^{6+}$ + He collisions \cite{Schulz03} 
(for a review see, e.g., \cite{Kouzakov12}). The effect of the NN interaction 
on the ionization depends on the sign of the projectile charge, therefore it is
expected to contribute to the  particle - antiparticle differences in FDCS.

Exhaustive reviews of the available theoretical differential studies of the
antiproton-induced ionization of hydrogen have been given in recent papers by 
Abdurakhmanov {\em et al.} \cite{Abdurak11} and Ciappina {\em et al.} 
\cite{Ciappina13}. In the followings we briefly summarize the models applied for
calculation of FDCS. In most of the works the authors compare the results of their
calculations with the predictions of the first Born approximations (FBA). Further,
fully quantum mechanical first-order perturbation approaches that include the NN
interaction are the continuum-distorted-wave-eikonal-initial-state (CDW-EIS) model 
of Voitkiv and Ullrich \cite{Voitkiv03}, that of Jones and Madison \cite{Jones02}, 
and the 3C model of Berakdar {\em et al.} \cite{Berakdar93}. Voitkiv  and Ullrich 
\cite{Voitkiv03} have also made calculations in the second-order Born approximation. 

Nonperturbative descriptions have also been applied in fully differential studies.
McGovern {\em et al.} \cite{McGovern09, McGovern10a, McGovern10b} 
worked out a model within the
framework of a time-dependent coupled pseudostate (CP) formalism. Although they 
used the straight-line approximation (SLA) for the projectile path, they could 
determine FDCS by establishing connection between the wave treatment of projectile
motion and the SLA method. In this way their model gives account of the NN 
interaction. The fully quantal, time-independent convergent close-coupling (CCC) 
model of Abdurakhmanov {\em et al.} \cite{Abdurak11} has been developed along the
lines of the CCC approach  to electron-atom scattering. The model is also based on 
use of pseudostates, and as a fully quantal theory, it implicitly considers the NN 
interaction. Recently Ciappina {\em et al.} \cite{Ciappina13} investigated the
differential properties of the antiproton-induced ionization within the framework 
of time-dependent close-coupling (TDCC) theory using SLA for the projectile
path. They employed a Fourier transform method in order to extract FDCS for a 
specific value of projectile momentum transfer, and included the NN interaction
into the model by a phase factor \cite{Gulyas08, Rodrig96} 
in the Fourier integral of the
transition amplitude over the impact parameter. For the sake of completeness we
mention that further investigations using semi-classical coupled-channel approaches 
\cite{Pons00a, Pons00b, Igarashi00, Sidky98} have also been reported in the
literature, but in these works only partially differential cross sections were
calculated or some special aspects of the antiproton - hydrogen collision were 
analyzed.

In this paper we report the results of an analysis carried out by the classical
trajectory Monte Carlo (CTMC) method. The motivation of the work was as follows.
From the comparison of the FDCS predicted by the above models it turned out
that there exist large discrepancies (more than factor of two) between the models,
particularly at low impact energies ($\le 200$ keV) \cite{Abdurak11, Ciappina13}. 
The reason of the discrepancies can be traced back to the approximations applied 
in the models. Most importantly, for the electronic wave function all the models 
use {\em single-center} expansion based on the target atom. The reasoning for
this approximation is that the antiproton has no bound states of electrons,
and therefore in lack of the electron capture channel there is no need to
include projectile-centered states in the expansion of wave function. However,
at low impact velocities a large distortion of the electron distribution
-- a strong reduction of the electron density near the antiproton --
is expected which cannot be represented by the one-center expansion, as it
was shown by Toshima \cite{Toshima01}. Probably, as a consequence of
the one-center approximation, in studies made with pseudostates the calculations 
were not repeated for protons, and therefore the analysis of one of the most 
interesting  characteristics of the future antiproton experiments, the 
particle - antiparticle difference in FDCS is missing in these studies.

CTMC provides an exact description of the full dynamics of the three-body break up 
process, albeit classically. It is known to reproduce the main features of the 
excitation, ionization and charge transfer processes in ion-atom collisions. It can be successfully used for calculations of differential cross sections (as an example, 
see Ref.\ \cite{Sarkadi10}). A further advantage of CTMC is that by analysis of the
calculated trajectories one can gain a deeper insight into the dynamics of the
collision processes. At the same time the model character of the method should be
emphasized: Because of the neglect of quantum mechanical effects CTMC has a limited
validity, in a number of applications it proved to be only a qualitative description.
For example, for proton on hydrogen collision CTMC underestimates the total ionization
cross section at 20 keV impact energy by more than a factor of two, and even at higher
proton energies it fails to reproduce the observed angular distribution of the
ejected electrons at backward angles \cite{Kerby95}.

We made the CTMC calculations at a relatively low impact energy of 30 keV where large
particle - antiparticle differences in FDCS are expected. Another reason for the 
choice of 30 keV was that at this energy FDCS calculations were performed in most 
of the quantum mechanical models, providing a basis for the comparison of the 
various approaches. To the best knowledge of the author, until the present work
the CTMC method has not been applied to study the full three-body dynamics of the
antiproton-induced ionization of the hydrogen atom, and even the number of such
studies for other collision systems involving positive ion projectiles 
is very scarce \cite{Fiol03, Schulz01}. At the same time, CTMC was applied in 
several works \cite{Meng93, Khayat99, Barna07, Tokesi09} to calculate partially
differential cross sections for the antiproton-induced ionization of the helium 
atom.

\section{Theoretical method}

The CTMC method is based on the numerical solution of the classical equations
of motion for a large number of trajectories of the interacting particles
under randomly chosen initial conditions \cite{Abrines66,Olson77}. 
The details of the used CTMC computer code are given in \cite{Sulik07}.
Briefly, it solves Newton's non-relativistic equations of motion for the
three particles (in atomic units):
\begin{equation}
m_i {{\rm d}^2 {\bf r}_i \over {\rm d}\, t^2} = 
    \sum_{j(\neq i)= 1}^3 Z_i Z_j {{\bf r}_i - {\bf r}_j 
    \over \;|{\bf r}_i - {\bf r}_j|^3}\, ,
    \; (i = 1, 2, 3)\,.
\label{eqno1}
\end{equation}
Here $m_i$, $Z_i$ and ${\bf r}_i$ are the masses, charges and 
position vectors of the particles, respectively. 
The randomly selected initial conditions were
the impact parameter and five further parameters defining the position and
velocity vector of the target electron moving on Kepler orbits. The ranges of
the latter parameters were constrained to give the binding energy of the
hydrogen atom, 0.5 a.u.. For the generation of the initial values of the
position and velocity coordinates of the electron from a set of uniformly
distributed variables we applied the general procedure suggested by Reinhold
and Falc\'on \cite{Reinhold86} for non-Coulombic systems which is equivalent
to the original Abrines and Percival's method \cite{Abrines66} in the case of
the Coulomb interaction.

The integration of the equations of motion was started at a large distance
(138 a.u.) between the incoming projectile and the hydrogen atom. 
After the collision the calculations were made in two steps. In the first step
the integration was continued until the internuclear distance $R = 138$ a.u.,
where the main reaction channels (excitation, ionization, and charge transfer
for proton impact) could be identified safely. In the second step only collision 
events leading to ionization were regarded. For the accurate determination
of the post-collisional effects on the electron emission 
\cite{Reinhold89,Sarkadi05}, in the second step the trajectories of the particles 
were calculated up to $R = 10^8$ a.u..

The fully differential cross section for ejection of
the electron with energy between $E_e$ and $E_e + {\rm d} E_e$ into solid angle 
${\rm d} \Omega_e$, and for scattering of the projectile into solid angle
${\rm d} \Omega_p$ is expressed classically as 
\begin{equation}
 {{\rm d}^3 \sigma \over {\rm d} E_e \, {\rm d} \Omega_e \, {\rm d} \Omega_p}
 = 2\pi\int_0^\infty b\,  
 {{\rm d}^3 P \over {\rm d} E_e \, {\rm d} \Omega_e\, {\rm d} \Omega_p} (b)\,
 {\rm d} b
 \, ,
\label{eqno2}
\end{equation}
where ${\rm d}^3 P / {\rm d} E_e \, {\rm d} \Omega_e\, {\rm d} \Omega_p$ is 
the fully differential ionization probability of the process, and $b$ is the impact
parameter. One can easily show that for large number $N$ of collision events
characterized by uniformly distributed $b$ values in the range 
$(0,b_{\rm max})$ the integral in (\ref{eqno2}) can be approximated by the 
following sum:
\begin{equation}
 \int_0^\infty b\,  
 {{\rm d}^3 P \over {\rm d} E_e \, {\rm d} \Omega_e\, {\rm d} \Omega_p} (b)\,
 {\rm d} b
 \approx {b_{\rm max}\, \Sigma_j b_j^{(i)} 
 \over N \Delta E _e\, \Delta \Omega_e \, \Delta \Omega_p}
 \, .
\label{eqno3}
\end{equation}
Here $b_j^{(i)}$ is the actual impact parameter at which the electron is emitted
into energy and solid angle window $\Delta E_e$ and $ \Delta \Omega_e$, and the
projectile is scattered into solid angle window $\Delta \Omega_p$. The solid
angles $\Delta \Omega_k$ ($k = e, p$) are determined by the minimum and maximum
values of the respective polar and azimuthal angles, $\theta_k$ and $\phi_k$:
\begin{eqnarray}
\Delta \Omega_k = 
\int_{\theta_{k}^{\;\rm min}}^{\theta_{k}^{\;\rm max}}
\int_{\phi_k^{\;\rm min}}^{\phi_k^{\;\rm max}}
\sin \theta_k\, {\rm d} \theta_k\, {\rm d} \phi_k = 
\nonumber
\\
(\cos \theta_{k}^{\;\rm min} - \cos \theta_{k}^{\;\rm max})
({\phi_k^{\;\rm max}} -{\phi_k^{\;\rm min}})
 \, .
\label{eqno4}
\end{eqnarray}
In our calculations we followed the history of $8\times 10^7$ ($1.6\times 10^8$)
collision events with $b_{\rm max} = 3.5$ (5) a.u. for antiproton (proton) impact.
We carried out two series of calculations: We repeated the computer runs for the
same collision events also without the NN interaction.

\section{Results and discussion}

For the total cross section of the ionization of the hydrogen atom by impact 
of 30 keV antiprotons CTMC resulted in $1.30\times 10^{-16}$ cm$^2$ that agrees 
with the measured  value of $(1.14 \pm 0.25)\times 10^{-16}$ cm$^2$ 
\cite{Knudsen95, Kirchner11} within  the experimental error. At the same time, the
corresponding value of $0.76\times 10^{-16}$ cm$^2$ for proton impact is smaller 
by 35\% than the  measured  value of $(1.18 \pm 0.026)\times 10^{-16}$ cm$^2$ 
\cite{Shah87}. We note that in lack of experimental data exactly at 30 keV, we
obtained the above cross section values by extrapolating and interpolating the
published data for the  antiproton and proton impact, respectively. As far as 
the NN interaction is concerned, it has a negligible ($< 1\%$) effect on the
calculated total cross sections for both projectiles.

From the results of the computer runs we derived FDCS values at electron energy 
$E_e = (5 \pm 1)$ eV and projectile scattering angle 
$\theta_p = (0.35 \pm 0.05$) mrad in the laboratory reference system.
The latter value corresponds to an average transverse momentum transfer 
$q_\bot = 0.7$ a.u.. We considered coplanar collision geometry, i.e., electron
emission events occurring in the collision plane were selected. The latter plane
is defined by the initial and final momentum of the projectile, ${\bf K}_i$ and
${\bf K}_f$, respectively. The condition of the coplanar electron emission 
was fulfilled by the choice $\phi_e - \phi_p = 0^\circ \pm 5^\circ$. The above
choice of the collisional parameters means that
for calculation of  FDCS in Eq.\ \ref{eqno3} we used $\Delta E_e = 2$ eV,
$\Delta \theta_p = 0.1$ mrad and $\Delta \phi_e = 10^\circ$. The azimuthally
isotropic scattering of the projectile was expressed by taking 
$\Delta \phi_p = 2\pi$. We mention here the main difficulty in calculation 
of FDCS by a Monte Carlo method, namely that the specification of the kinematical
parameters of the collision by sufficiently narrow windows strongly reduces the
number of the regarded ionization events, and to achieve a reasonable counting
statistics one needs to follow the history of very large number of collisions.

\begin{figure}
\includegraphics[angle=0,width=0.45\textwidth]{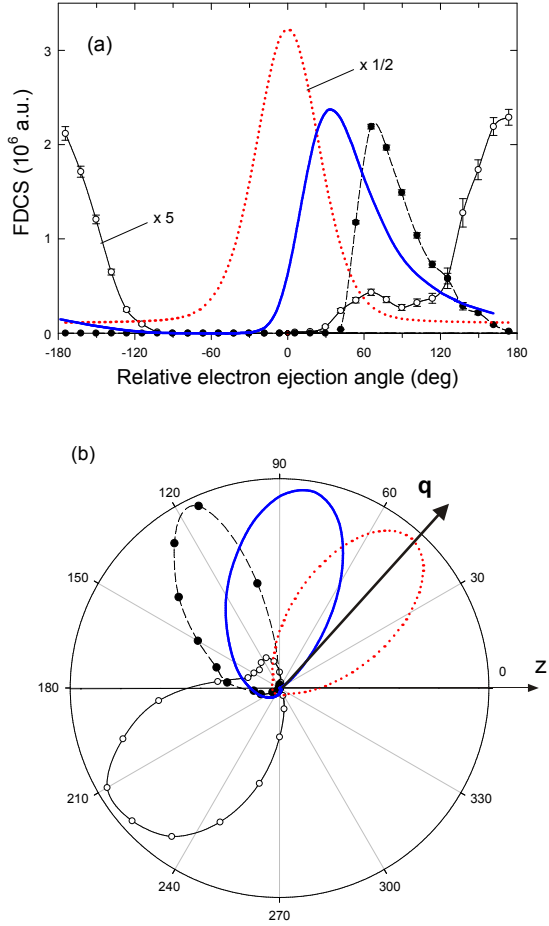}
%\vspace*{-0.3cm} 
\caption{(Color online)
FDCS for ionization of the hydrogen atom by impact of 30 keV antiprotons
in the scattering plane. The energy of the ejected electron is 5 eV, 
the scattering angle of the projectile is 0.35 mrad. Open circles,
CTMC including the NN interaction; solid circles, CTMC neglecting the NN
interaction; thick solid line (blue), CDW neglecting the NN interaction;
dotted line (red), FBA. (a) The angular distribution of the electron
as a function of the difference between the electron ejection angle 
$\theta_e$ and the direction of the momentum transfer vector $\theta_q$.
(b) Polar diagram of the electron electron emission as a function of the
electron ejection angle $\theta_e$. The distributions in the polar diagram 
are normalized at their maximum values. The arrow labeled by q shows the
direction of the momentum transfer. The $z$ axis defines the direction of 
the incoming projectile beam. The thin solid and dashed lines through the 
CTMC results are only to guide the eye.
}
\label{fig1}
\end{figure}

\begin{figure}
\includegraphics[angle=0,width=0.45\textwidth]{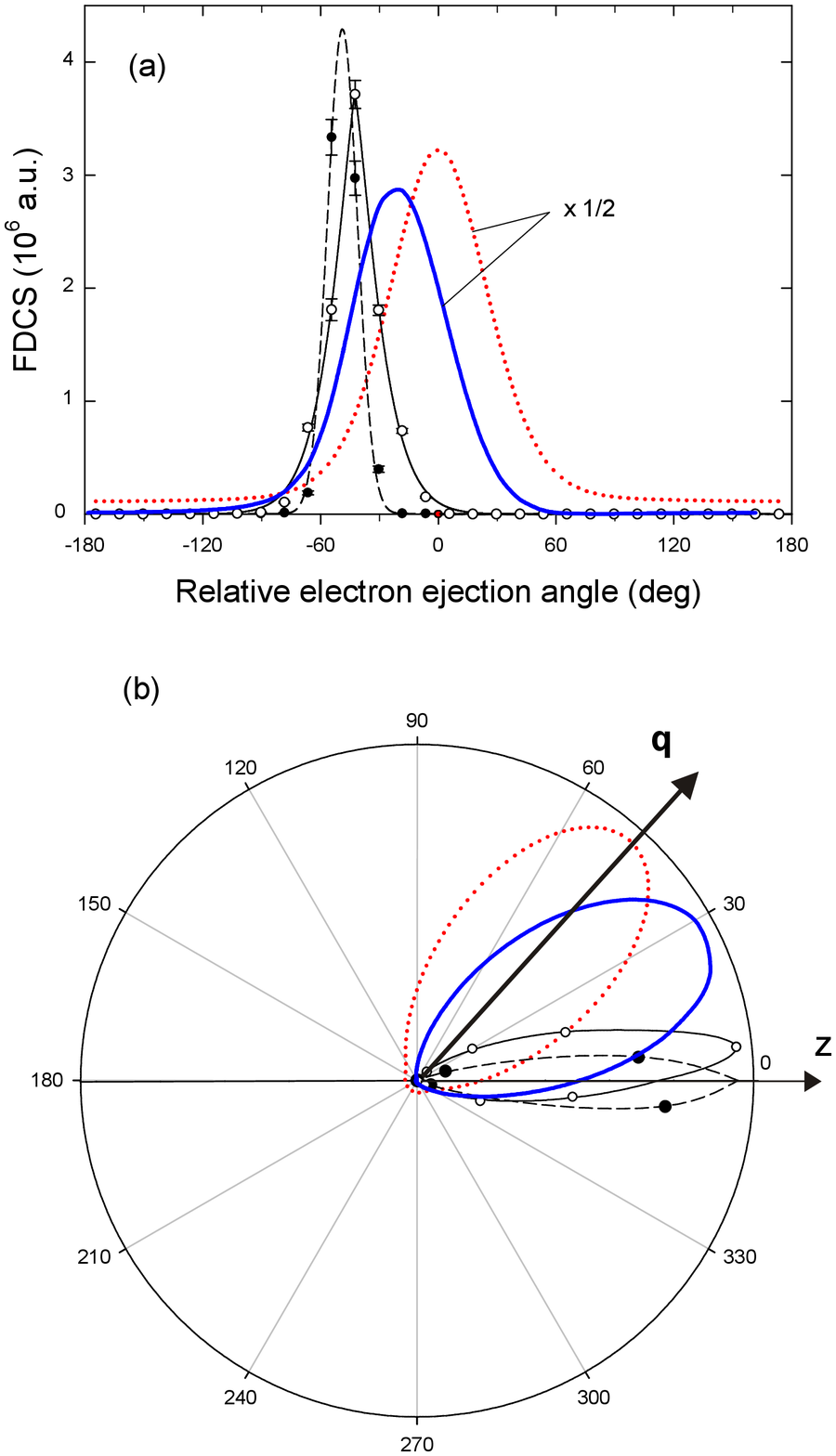}
\vspace*{0.2cm} 
\caption{(Color online)
The same as Fig.\ \ref{fig1} but for proton impact.
} 
\label{fig2}
\end{figure}

The results of the calculations for antiproton and proton impact are presented in 
Fig.\ \ref{fig1} and Fig.\ \ref{fig2}, respectively. In the figures we plotted also 
the prediction of FBA. The latter cross section can be expressed analytically 
(see, e.g., \cite{McGovern09}). In the laboratory frame (in atomic units):
\begin{eqnarray}
 {{\rm d}^3 \sigma \over {\rm d} E_e \, {\rm d} \Omega_e \, {\rm d} \Omega_p}
 = {256\,Z_p^2 \, m_p^2 \, v_f \over v_0 q^2 \pi [1 - \exp (-2\pi/\kappa)]} \times
\nonumber
\\
\nonumber
\\
{\exp[-{2 \over \kappa } \arctan ({2\kappa \over 1 + q^2 - \kappa^2})] 
\over (1 + q^2 - \kappa^2)^2 + 4\kappa^2} \times
\nonumber
\\
\nonumber
\\
{q^2 - 2{\bm\kappa} {\bf q} + {(\kappa^2 +1) \over \kappa^2 q^2} 
({\bm\kappa} {\bf q})^2 \over (1 +q^2 + \kappa^2 - 2{\bm\kappa} {\bf q})^4}
 \, .
\label{eqno5}
\end{eqnarray}
We note that the sign of the terms $2{\bm\kappa} {\bf q}$ 
in Eq.\ (\ref{eqno5}) differs from that in Eq. (77) of Ref.\ \cite{McGovern09},
but agrees with that in Eq.\ (7.2.31) of Ref.\ \cite{McDowell70}.

In a naive view of the ionization the electron is expected to fly out from the
atom in the direction of the transferred momentum due to the dominant 
electron-projectile interaction, i.e., the angular distribution of the electron
is expected to be peaked at $\theta_e = \theta_q$ (at the present
collisional  parameters $\theta_q = 48^\circ$). This explains why FDCS is
plotted against the relative electron emission angle  $\theta_e - \theta_q$
in Figs.\ \ref{fig1}a and \ref{fig2}a. In panel (b) of the figures we plotted 
also the dependence of FDCS on $\theta_e$ in form of polar diagram. In the 
latter diagram the emphasis was put on the directional information, therefore 
the distributions were normalized at their maximum values.

As is seen from the figures, FBA predicts forward electron 
emission in the direction of the momentum transfer,
in accordance with the aforementioned expectation. This can be understood 
considering that FBA accounts only for the projectile-electron interaction.
Furthermore, FBA yields equal FDCS for antproton and proton impact because of
the $Z_p^2$ dependence on the projectile charge.

The FBA peak in Figs.\ \ref{fig1} and \ref{fig2} is a result
of a direct momentum transfer in binary collision between the projectile and the
electron, therefore it is called as "binary peak". At suitable collision conditions
(higher collision velocity and lower projectile scattering angle) a second 
structure (called "recoil peak") also appears in the angular distribution.
It has maximum in the direction of $-\bf q$, and it is interpreted as a
double scattering process: First the electron is ejected via binary interaction
with the projectile with momentum $\bf q$, then in its way out of the atom
it backscatters elastically from the target nucleus (see, e.g., 
Ref.\ \cite{Schulz03}). 

For both projectiles the present CTMC calculations resulted in electron emission 
into completely different directions than that predicted by FBA. First we
discuss the case of antiproton impact. Even without the NN interaction the 
obtained electron distribution is peaked at a backward angle, at
 $\theta_e \approx 120^\circ$ (see Fig.\ \ref{fig1}b). 
Interestingly, the shape of this latter
distribution is similar to that of FBA: The two peaks have about the same width, but
the distribution predicted by CTMC shows some asymmetry.
This is in a qualitative agreement with  the TDCC results of Ciappina et al. 
\cite{Ciappina13} obtained without inclusion  of the NN interaction. Concerning the
peak intensities, the CTMC result is smaller  by a factor of 3 than that of FBA. 
The inclusion of the NN interaction led to a dramatic effect: The FDCS is further
reduced by a factor of 5, and the angular distribution completely changed. 
In this case the electrons are emitted at even larger backward angles. The
distribution has maximum at $\theta_e \approx 220^\circ$, but a smaller peak is also
visible at $\theta_e \approx 110^\circ$. Similar double-peak structure has been observed
in quantum mechanical calculations \cite{Voitkiv03, Abdurak11, Ciappina13}.
In the latter works the smaller and the larger peak were identified as the
binary and the recoil peak. In the followings we will refer the two peaks using
these notations.

A very different result was obtained for proton impact. Our both calculations
without and with inclusion of the NN interaction show an opposite shift of the
binary peak as compared to antiproton impact: The electrons are emitted at
small angles in forward direction. The widths of the distributions are much narrower
than that predicted by FBA. This indicates the presence of a strong two-center effect.
The intensities of the peaks are smaller than that predicted by FBA, but the difference 
is smaller for protons than for antiprotons.

In Figs.\ \ref{fig1} and \ref{fig2} we plotted also FDCS data obtained from
CDW-EIS calculations \cite{Gulyas08} without considering the NN interaction.
CDW-EIS is also a perturbation theory as FBA, but unlike FBA it accounts for the
distortion of the electronic states in the presence of the projectile. Therefore,
CDW-EIS is expected to provide FDCS data that are closer to the CTMC results.
Indeed, for antiproton impact of the binary peak  predicted by CDW-EIS is very 
similar to that obtained from CTMC, regarding both its the intensity and shape.
At the same time, CDW-EIS predicts a smaller shift of the peak from the 
direction of ${\bf q}$ than CTMC. The widths the peaks also differ slightly, the 
CDW-EIS peak is broader. For proton impact CDW-EIS predicts much smaller
two-center effect than CTMC, the CDW-EIS peak does not show the strong narrowing
effect observed for CTMC. 

\begin{figure}
    \hspace*{-1cm}
{\includegraphics[angle=0,width=0.5\textwidth]
         {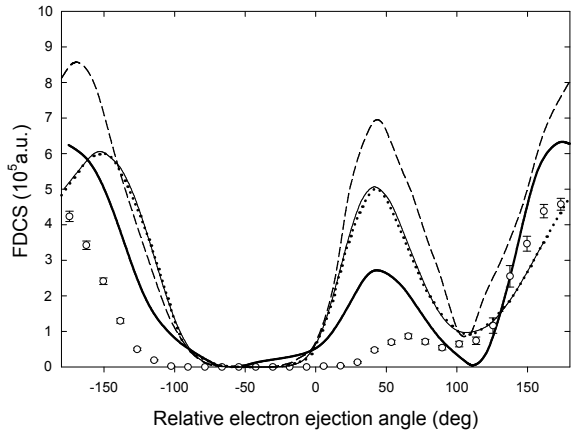}} 
 \caption{
Comparison of FDCS for antiproton impact obtained by the present CTMC
calculations  with the results of quantum mechanical models . All the calculations 
were made with inclusion of the NN interaction. Open circles with error bar, CTMC;
thick solid line, TDCC \cite{Ciappina13}; dashed line, CCC \cite{Abdurak11}; dotted
line, CDW-EIS \cite{Voitkiv03}; thin solid line, CDW-EIS calculated in the present 
work.
 }
\label{fig3}
\end{figure}

\begin{figure}
\includegraphics[angle=0,width=0.45\textwidth]{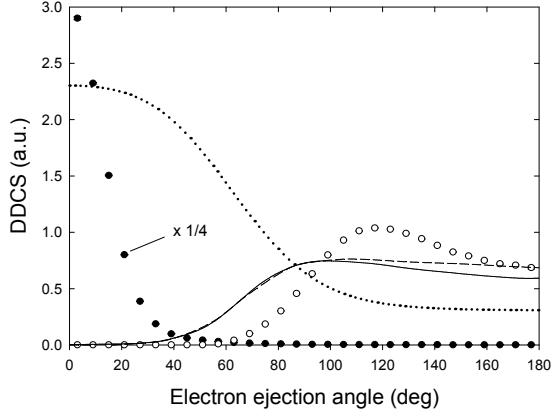}
\caption{
DDCS for ejection of electrons of energy 5 eV as a function of the emission
angle. The dotted line denotes FBA. The present CTMC results are shown
by open and solid circles for antiproton and proton impact, respectively.
Quantum mechanical models for antiproton impact: solid line, CCC \cite{Abdurak11};
dashed line, CP \cite{McGovern09}.
}
\label{fig4}
\end{figure}

In Figs.\ \ref{fig3} and \ref{fig4} we compare the present FDCS and DDCS results 
with those of quantum mechanical calculations.  In Fig.\ \ref{fig3} the quantum
mechanical models used in the comparison are the TDCC theory of  Ciappina 
{\em et al.} \cite{Ciappina13}, the CCC approach of Abdurakhmanov 
{\em et al.} \cite{Abdurak11}, and the CDW-EIS model of Voitkiv and Ullrich 
\cite{Voitkiv03}. The FDCS data of the latter model were taken from 
Ref.\ \cite{Abdurak11}, as well as we made independent CDW-EIS  calculations 
also in the present work. 

We note that the present CTMC data were evaluated in the 
laboratory reference system. At the same time, the published FDCS results of the 
above models were expressed in the {\em relative} coordinate system. To convert
the latter data to the laboratory system we multiplied them with the factor
$(m_p/\mu)^2$ (see, e.g., \cite{McGovern10b}), where $\mu_p$ is the reduced mass 
of the projectile:
\begin{equation}
\mu_p = {m_p \, m_H \over m_p + m_H}
 \, .
\nonumber
\end{equation}
Here $m_p$ and $m_H$ are the mass of the projectile and that of the hydrogen 
atom, respectively. For proton (antiproton) on hydrogen scattering to a good
approximation $(m_p/\mu)^2 \approx 4$. Unlike FDCS, the DDCS values are the same 
in the two reference systems (due to the integration over $\theta_p$), therefore
we did not need to correct the DDCS data.

From Fig.\ \ref{fig3} we can establish only qualitative agreement between CTMC
and the quantum mechanical models. The disagreement is particularly large for the
binary peak concerning both its intensity and position. While all the three quantum
mechanical models predict a peak position of about $45^\circ$, according to CTMC the
peak appears at about $65^\circ$. CTMC predicts a greatly suppressed binary peak.
For the recoil peak a better agreement is observed. In the latter case a striking
feature of the CTMC results is the narrower peak width compared to other
theories. 

Concerning the greatly suppressed binary peak predicted by CTMC, we note that the
description of this peak seems to be very sensitive to the the applied theoretical
approach: Even for the quantum mechanical models the peak maximum
varies by a factor of more than two. Furthermore, we note that the coupled 
pseudostate (CP) calculations of McGovern {\em et al.} \cite{McGovern10b} carried
out under identical collision conditions with the present work resulted in
also a greatly suppressed binary peak relative to the recoil peak (see the 3D plot
of FDCS in Fig.\ 1 of Ref.\ \cite{McGovern10b}). 

There may be several reasons of the discrepancies between CTMC and the quantum
mechanical models. As is seen from Fig.\ \ref{fig1}, the inclusion of the NN 
interaction has a profound effect on FDCS, therefore its approximate treatment
may introduce uncertainties into the calculations. 

As it is discussed in the Introduction, the coupled-states descriptions (TDCC, CCC)
are based on one-center expansion of the electronic wave function. This approximation
is questionable at low impact energy \cite{Toshima01}, thus the neglect of the 
two-center effects may be a further reason of the discrepancies. We note that in 
CDW-EIS the distortion factors applied at the initial- and final-state wave function
give account of the two-center effect. However, CDW-EIS is a perturbation theory, its
use is justified at high impact energy.

As far as CTMC is concerned, it remains a question how far FDCS is affected by the
neglect of the quantum mechanical effects. Anyhow, CTMC seems to be suitable for
the differential characterization of the antiproton-induced ionization of the 
hydrogen atom, and may contribute in this way to a deeper understanding of the 
process.

In Fig.\ \ref{fig4} the present DDCS results are compared with those of the
the CCC approach of  Abdurakhmanov {\em et al.} \cite{Abdurak11} and the CP model
of McGovern {\em et al.} \cite{McGovern09} as a function of the electron emission 
angle. The energy of the electron is 5 eV. The CTMC data are plotted for both
antiproton and proton impact, and demonstrate well the expected large 
particle-antiparticle difference.  CTMC predicts dominant electron emission at 
backward directions for antiproton impact, in qualitative agreement with the  
quantum mechanical models.

To investigate the role of the NN interaction in the antiproton-induced ionization,
we analyzed particle trajectories at various collision conditions. 
As a great surprise, practically no difference was observed in the electron 
trajectories when the NN interaction was turned on and off. This is in contrast to 
the previous explanation of the NN interaction effect given by 
Abdurakhmanov {\em et al.} \cite{Abdurak11}, who assumed an interference effect that 
takes place between the interactions of the target electron and proton with the
outgoing antiproton. According to the authors, the outgoing scattered antiproton is
decelerated in the attractive field of the target nucleus, resulting in a stronger
final-state interaction between the antiproton and the electron. This leads to the
polarization of the target electron cloud and a shift of the electron density away 
from the projectile path.

The insensitivity of the electron trajectories on the NN interaction observed in the
present work indicates that the effect assumed by Abdurakhmanov {\em et al.} is
probably very small at the collision energies regarded also by the authors 
($\geq 30$ keV). Then the question is: How can one explain the drastic change of
FCDS seen in Fig.\ \ref{fig1} when the NN interaction is turned on?

The answer was found by analyzing the trajectory of the target nucleus. We found
that it changed in a large extent when the NN  interaction was turned on. The change 
is caused by the momentum transferred by the projectile to the target nucleus in the 
NN scattering. As a result, the total momentum transferred to the whole atom is also 
changed which leads to the rearrangement of the  collision events and to a modified 
angular distribution.

The finding that the nucleon-nucleon scattering has practically no effect on the motion 
of the electron is understandable considering the very small scattering angle and 
the negligible change in collision velocity, as well as the  length scale difference of 
three orders of magnitude between the motion of the electron and that of the target 
nucleus. 
\begin{figure}
\includegraphics[angle=0,width=0.45\textwidth]{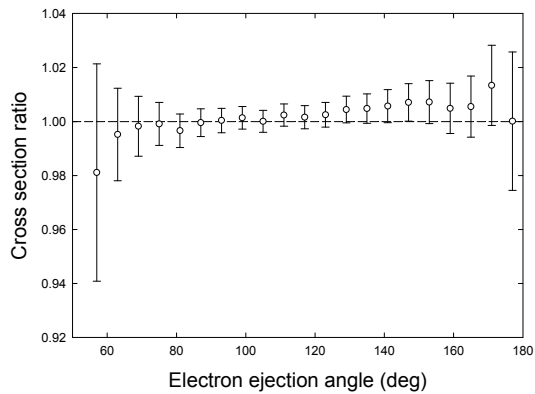}
\caption{
Ratio of DDCS values calculated with inclusion of the NN interaction to those calculated 
without it. The energy of the ejected electron is 5 eV.
}
\label{fig5}
\end{figure}
The rigidity of the angular distribution of the electron on the NN interaction is well
reflected by the ratio of DDCS values for ejection of electrons of 5 eV calculated with
and without the NN interaction. The ratio is plotted in Fig.\ \ref{fig5} for electron 
ejection angles $\theta_e > 50^\circ$ at which DDCS takes appreciable values. 
Although systematical deviations from unity can be observed at smaller and larger angles, 
the effect is small ($< 2\%$) and within the error of the calculations.

Our finding that the relative motion between the electron and the nuclei is coupled
very weakly with that between the nuclei indicates that two motions can be treated 
independently. This led us to show that the NN interaction can be be included in the
calculations in the form of the following simple correction procedure. Let us denote
the additional momentum transfer vector due to the NN scattering by ${\bf q}^{\rm NN}$.
For small scattering angles the longitudinal component of ${\bf q}^{\rm NN}$ can be
neglected, and the transversal component is given as
\begin{equation}
q^{\rm NN}_\bot \approx k_i \, \theta_p^{\rm NN}
 \, .
\label{eqno6}
\end{equation}
$\theta_p^{\rm NN}$ is the NN two-body scattering angle that can be obtained
from the relationship
\begin{equation}
\theta_p^{\rm NN} = 2 \arctan \left({b \over a} \right)
 \, ,
\label{eqno7}
\end{equation}
where $a = Z_p Z_t / 2E_p$ is the half distance of closest approach ($E_p$: the energy
of the projectile). 

\begin{figure}
\includegraphics[angle=0,width=0.45\textwidth]{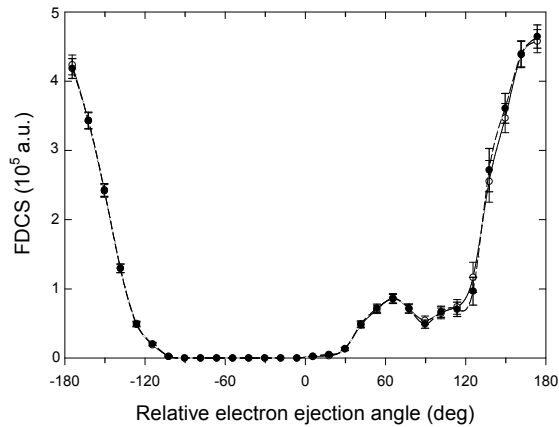}
\caption{
Comparison of FDCS values obtained by approximate treatment of the NN interaction
effect (solid circles) with the exact results (open circles).
}
\label{fig6}
\end{figure}

The correction procedure is simply the replacement of the momentum transfer vector
$\bf q$ by the vector $\bf q + {\bf q}^{\rm NN}$ for all the collision events that 
were calculated without the NN interaction. The FDCS data derived from the modified
collision events are compared with those obtained with the "exact" treatment
of the NN effect in Fig.\ \ref{fig6}. We may conclude from the figure that the 
correction procedure is excellent, thus proving the weak coupling between the 
electron-nuclei and the nucleon-nucleon relative motion. We note that the success of
the presented approximate treatment of the NN effect gives a strong support to the
procedure applied by Schulz {\em et al.} \cite{Schulz07} in the analysis of their
experimental FDCS results obtained for ionization in 100 MeV/amu C$^{6+}$ + He
collisions. The latter authors used the Monte Carlo event generator (MCEG) method
in FBA to account for the additional momentum transfer due to the elastics 
scattering of the projectile ion on the target nucleus. The application of MCEG was
necessary, because such a correction can be made only event-by-event, in a way as it
was done in our present CTMC investigation.

\section{Conclusions}

We investigated the three-body dynamics of the ionization of the atomic hydrogen 
induced by antiprotons. To this end, we calculated fully differential cross 
sections by applying the CTMC method. The calculations were made at relatively
low impact energy of 30 keV where large deviations from the predictions of the
first Born approximation are expected. The kinematical parameters (electron energy,
projectile scattering angle) were chosen to be those of quantum mechanical 
investigations of the process available in the literature. The calculations
made also for proton impact under the same collision conditions revealed large
particle-antiparticle differences in FDCS. Comparing the CTMC results with
the predictions of quantum mechanical models (CCC, TDCC, CDW-EIS) we concluded 
that the classical mechanical description can reproduce the main features
of the antiproton-induced ionization of the hydrogen atom, and thereby it
helps the deeper understanding of the process. We analyzed the possible
reasons of the observed discrepancies between CTMC and the quantum mechanical 
models: The approximate treatment of the NN interaction and the use of the 
one-center expansion of the electronic wave function in the quantum mechanical
descriptions on one side, and the neglect of quantum effects in CTMC on the other 
side. 

To clarify the role of the NN interaction in the ionization, we examined 
individual particle trajectories. We established that the relative motion between 
the electron and the nuclei is coupled very weakly with that between the nuclei, 
consequently the two motions can be treated independently. This was convincingly
proved by a calculation in which the additional momentum transfer due to the elastics 
scattering of the projectile on the target nucleus was taken into account by a
simple correction procedure for collision events obtained without inclusion
of the NN interaction.

\section{Acknowledgments}

This work was supported by the National Scientific Research Foundation (OTKA, 
Grant No.\ K109440) and the National Information Infrastructure Program (NIIF).

\end{document}